\documentstyle[multicol,prl,epsf,aps]{revtex}
\begin{document}
\tightenlines

 

\title{Self-interacting dark matter from an SU(3)$_L$$\otimes$U(1)$_N$ electroweak model}
\author{D. Fregolente and M. D. Tonasse}  
\address{Instituto Tecnol\'ogico de Aeron\'autica, Centro T\'ecnico  
Aeroespacial, Pra\c ca Marechal do Ar Eduardo Gomes 50, 12228-900 S\~ao Jos\'e 
dos Campos, SP, Brazil}  
\maketitle    
\begin{abstract}  
Recently it has been shown that models which consider self-interacting particles as dark matter candidates can be able to account with practically all the discrepancies between N-body simulations and astronomical observations of the galactic structure. In the present work we show that the 3-3-1 electroweak model can provide a realistic candidate to this kind of dark matter. This dark matter particle is not arbitrarily imposed and no new symmetry is needed to stabilize it.
\end{abstract}


\begin{multicols}{2}
\narrowtext

For long time physicists, astronomers and cosmologists have dedicated a much effort towards improving our knowledge about the structure and the content of the Universe. Although new observational techniques were able to implement important advances at the end of the last century, several fundamental problems remain unaswered. In particular, it is a curious fact that we ignore the nature of about 90$\%$ of all matter. The presence of this {\it dark matter} in the Universe is claimed for both observational and theoretical predictions and it is the object of much speculation.\par
From the observational side, the presence of dark matter receives support from measurements of the rotational velocity of stars relative to galactic centers, galactic clusters dynamics, X-ray detection from gas in galactic clusters, gravitational lenses and flux velocities on scales larger than cluster scales. Theoretical arguments come from inflationary theory \cite{KM01}.\par
It is a well-established fact that the plausible candidates to this matter are elementary particles. The standard model offers no options for dark matter. However, its extensions present a proliferation of exotic particles and so, it is possible that some of them have properties that qualify them as good dark matter candidates. A small fraction of the dark matter is known to be of baryonic origin \cite{KM01} and, since we now know that neutrinos are massive  \cite{Aea99}, they also contribute with a small fraction. Recently, astronomical observations suggest that about 70$\%$ of the total energy of the Universe can be associated to the cosmological constant \cite{Nea01}. Thus, the contribution of an exotic particle to dark matter would be about 30$\%$. \par
Until a few years ago, the more satisfactory cosmological scenarios were those ones composed of ordinary matter, cold dark matter and a contribution associated with the cosmological constant. To be consistent with inflationary cosmology, the spectrum of density fluctuations would be nearly scale-invariant and adiabatic. However, in recent years it has been pointed out that the conventional models of collisionless cold dark matter lead to problems with regard to galactic structures. They were only able to fit the observations on large scales ($\gg 1$ Mpc). Also, $N$-body simulations in these models result in a central singularity of the galactic halos \cite{GM00} with a large number of sub-halos \cite{MG99}, which are in conflict with astronomical observations. A number of other inconsistencies, which we will not describe here, are discussed in Refs. \cite{DS01,BK00}. Thus, the cold dark matter model is not able to explain observations on scales smaller than a few Mpc. \par 
However, it has recently been shown that an elegant way to avoid these problems is to assume the so called {\it self-interacting dark matter} \cite{SS00}. One should notice that, in spite of all, self-interacting models lead to spherical halo centers in clusters, which is not in agreement with ellipsoidal centers indicated by strong gravitational lensing observations \cite{Yea00} and by Chandra observations \cite{BJ02}. Chandra observations also appears to find inconsistency with the flattened core DM profiles predicted for self interacting dark matter \cite{BJ02,LB02}. Although $N$-body simulations and astronomical observations are not conclusive concerning their predictions \cite{DS01,KK00} yet, self-interacting dark matter models are self-motived as alternative models. The key property of this kind of matter is that, although its annihilation cross-section is suppressed, its scattering cross section is enhanced. It allows efficient heat transport from the dark matter halo to the galactic center. The effect of this heat transfer is an expansion and a heating of the galactic nuclei that is crucial to account for the observed galactic properties. \par
Several authors have proposed models in which a specific scalar singlet that satisfies the self-interacting dark matter properties is introduced in standard model in an {\it ad hoc} way \cite{MC02,BP01}. To be stable this scalar can not interact strongly with the standard model particles and it is guaranteed by introduction of an extra symmetry (usually an U(1)). \par
In this Letter, for the first time, we examine the possibility of the self-interacting dark matter proposal being realized in a realistic gauge model, which was proposed with an independent motivation. Our argumentation is in the context of the so-called 3-3-1 electroweak gauge model. In this model the weak and electromagnetic interactions are described through a gauge theory based on the SU(3)$_L\otimes$U(1)$_N$ symmetry group \cite{FH93}. The most interesting feature of the model is that the anomaly cancellations occur only when the three-fermion generations are considered together and not family by family as in the standard model. This implies that the number of families must be a multiple of the color number. In addition, if we take into account that the asymptotic freedom condition of QCD imposes that the number of generations is less than five, the model predicts only three families. Another interesting feature is that the model can manifest itself at a relatively low mass scale through processes that violate individual lepton number. Therefore, there is a reasonable set of motivations for searching typical 3-3-1 processes in the next generation of colliders \cite{CQ99}. \par
There are several versions of the 3-3-1 model and all of them preserve the essential features discussed above \cite{FH93,TO02,PT93}. The original version of Refs. \cite{FH93} has a scalar sector with three Higgs triplets and one sextet. However, since we work here mainly with the scalar sector, it is convenient to take a version of the model with a smaller number of Higgs fields. Therefore, for the sake of simplicity, we will hereafter keep in mind the model of Ref. \cite{PT93}, which requires only three Higgs triplets to give the symmetry breaking and correct mass spectrum.\par
Let us summarize the most relevant points of the model of Ref. \cite{PT93}. The left-handed lepton and quark fields form the SU(3)$_L$ triplets,  $\psi_{aL} = \pmatrix{\nu_a & \ell_a & P_a}^{\tt T}_L \sim \left({\bf 3}, 0\right)$,  $Q_{1L} = \pmatrix{u_1 & d_1 & J_1}^{\tt T}_L \sim \left({\bf 3}, 2/3\right)$ and $Q_{\alpha L} = \pmatrix{J_\alpha & u_\alpha & d_\alpha}^{\tt T}_L \sim \left({\bf 3}^*, -1/3\right)$. Throughout this work we will use the convention that Latin indices run over 1, 2, and 3, while Greek ones are 2 and 3. $\ell_a = e, \mu, \tau$ and $P_a$ are exotic leptons. Each charged left-handed fermion field has its right-handed counterpart transforming as a singlet of the SU(3)$_L$ group.\par
In the gauge sector the model predicts, in addition to the standard $W^\pm$ and $Z^0$, the extra $V^\pm$, $U^{\pm\pm}$ and ${Z^\prime}^0$ gauge bosons.\par
The fermion and gauge boson masses are generated through the three Higgs scalar triplets $\eta = \pmatrix{\eta^0 & \eta_1^- & \eta^+_2}^{\tt T} \sim \left({\bf 3}, 0\right)$,  $\rho = \pmatrix{\rho^+ & \rho^0 & \rho^{++}}^{\tt T} \sim \left({\bf 3}, 1\right)$, $\chi = \pmatrix{ \chi^- & \chi^{--} & \chi^0}^{\tt T} \sim \left({\bf 3}, -1\right)$. The neutral scalar fields develop vacuum expectation values $\langle\eta^0\rangle = v$, $\langle\rho^0\rangle = u$, and $\langle\chi^0\rangle = w$, with $v^2 + u^2 = v_W^2 = 246^2 {\mbox{ GeV}}^2$. The pattern of symmetry breaking is SU(3)$_L\-\otimes$\-U(1)$_N$\-$\stackrel{\langle\chi\rangle}{\longmapsto}\- $SU(2)$_L\otimes$\-U(1)$_Y$\-$\stackrel{\langle\rho, \eta\rangle}{\longmapsto}$\-U(1)$_{{\rm em}}$. So, it is reasonable to expect $w \gg v, u$.\par
The most {\it economical}, gauge invariant and renormalizable Higgs potential is 
\end{multicols}
\hspace{-0.5cm}
\rule{8.7cm}{0.1mm}\rule{0.1mm}{2mm}
\widetext
\begin{eqnarray}
V\left(\eta, \rho, \chi\right) & = & \mu_1^2\eta^\dagger\eta + \mu_2^2\rho^\dagger\rho + \mu_3^2\chi^\dagger\chi + a_1\left(\eta^\dagger\eta\right)^2 + a_2\left(\rho^\dagger\rho\right)^2 + a_3\left(\chi^\dagger\chi\right)^2 + \left(\eta^\dagger\eta\right)\left(a_4\rho^\dagger\rho + a_5\chi^\dagger\chi\right) + \cr && a_6\left(\rho^\dagger\rho\right)\left(\chi^\dagger\chi\right) + a_7\left(\rho^\dagger\eta\right)\left(\eta^\dagger\rho\right) + a_8\left(\chi^\dagger\eta\right)\left(\eta^\dagger\chi\right) + a_9\left(\rho^\dagger\chi\right)\left(\chi^\dagger\rho\right) + \frac{1}{2}\left(f\epsilon^{ijk}\eta_i\rho_j\chi_k + {\mbox{H. c.}}\right).
\label{pot}\end{eqnarray} 
\hspace{9.1cm}
\rule{-2mm}{0.1mm}\rule{8.7cm}{0.1mm}
\begin{multicols}{2}
\narrowtext\noindent
Here the $\mu$'s and $f$ are coupling constants with dimension of mass with $a_3 < 0$ and $f < 0$ from the positivity of the scalar masses \cite{TO96}.\par
Symmetry breaking is initiated when the scalar neutral fields are shifted as $\varphi = v_\varphi + \xi_\varphi + i\zeta_\varphi$, with $\varphi$ $=$  $\eta^0$, $\rho^0$, $\chi^0$. The details of the physical spectrum of the neutral scalar sector are crucial for our results. It is given in Refs. \cite{TO96} and we summarize them here. Firstly we notice that real part of the shifted fields leads to the three massive physical scalar fields $H_1^0$, $H_2^0$, $H_3^0$ defined by 
\begin{equation}
\pmatrix{\xi_\eta \cr \xi_\rho} \approx \frac{1}{v_W}\pmatrix{v & u \cr u & -v\cr}\pmatrix{H^0_1 \cr H^0_2 \cr}, \qquad \xi_\chi \approx H_3^0,
\label{3H0}
\end{equation}
where we are using $w \gg v, u$. The scalar $H^0_1$ is the one that we can identify with the standard  model Higgs, since its squared mass,
\begin{equation}
m_1^2 \approx 4\frac{a_2u^4 - a_1v^4}{v^2 - u^2},
\label{h1}
\end{equation}
carries no any feature from the 3-3-1 breakdown to the standard model. On the other hand $H^0_3$, with squared mass
\begin{equation}
m_3^2 \approx -4a_3w^2,
\label{h3}
\end{equation}
is a typical 3-3-1 scalar. So, there is no any massless Goldstone boson rising from the real part of the neutral sector. On the other hand, from the imaginary part we have two Goldstone and one massive physical state $h^0$ with eigenstate
\begin{equation}
\zeta_\chi \approx h^0
\label{zetac}
\end{equation}
and squared mass
\begin{equation}
m_h^2 = -f\frac{v_W^2w^2 + v^2u^2}{vuw}.
\label{mh}
\end{equation}
It is important to notice that $\zeta_\eta$ and $\zeta_\rho$ are pure massless Goldstone states. The approximation in Eqs. (\ref{3H0}), (\ref{h1}), (\ref{h3}) and (\ref{zetac}), is valid for $w \gg v, u$. This condition leads to relations among the parameters of the scalar potential (\ref{pot}). One of them, which enters in the $H_1^0h^0h^0$ interaction, is
\begin{equation}
a_5v^2 + 2a_6u^2 \approx -\frac{vu}{2}
\label{ll}
\end{equation}
(see Ref. \cite{TO96}).\par
The covariant derivative is given by
\begin{equation}
{\cal D}_\mu\phi_i = \partial_\mu\phi_i + i\frac{g}{2}\left(\vec {W}_\mu.\vec{\lambda}\right)^j_i\phi_j + ig^\prime N_{\phi_i}B_\mu\phi_i, 
\label{deri}
\end{equation}
where $\vec W_\mu$ and $B_\mu$ are field tensors of SU(3) and U(1) gauge groups, respectively. The $\lambda$'s are Gell-Mann matrices, $g$ and $g^\prime$ are coupling constants associated with SU(3) and U(1), respectively and $\phi_i =$ $\eta$, $\rho$ and $\chi$, with $N_\eta = 0$, $N_\rho = 1$ and $N_\chi = -1$. \par  
We must consider also the matter coupling through the scalar fields. In the model of Ref. \cite{PT93}, the full Yukawa Lagrangians that must be considered are 
\end{multicols}
\hspace{-0.5cm}
\rule{8.7cm}{0.1mm}\rule{0.1mm}{2mm}
\widetext
\begin{mathletters}
\begin{eqnarray}
{\cal L}_\ell & = & -\sum_{ab}\left(\frac{1}{2}\epsilon^{ijk}G^{\left(\nu\right)}_{ab}\overline{{\psi_{aiL}}^C}\psi_{bjL}\eta_k + G^{\left(\ell\right)}_{ab}\overline{\psi}_{aL}\ell^-_{bR}\rho - G^{\left(P\right)}_{ab}\overline{\psi}_{aL}P^+_{bR}\chi\right)
\label{yukl} + {\mbox{H. c.}}, \\
{\cal L}_Q & = & \overline{Q}_{1L}\sum_b\left(G^{\left(U\right)}_{1b}U_{bR}\eta + G^{\left(D\right)}_{1b}D_{bR}\rho + G^{\left(J\right)}J_{1R}\chi\right) + \cr
&& \sum_\alpha\overline{Q}_{\alpha L}\left(F^{\left(U\right)}_{\alpha b}U_{bR}\rho^* + F^{\left(D\right)}_{\alpha b}D_{bR}\eta^* + \sum_{\beta}F^{\left(J\right)}_{\alpha\beta}J_{\beta R}\chi^*\right) + \mbox{H. c.},
\label{yukq}\end{eqnarray}\label{yukt}\end{mathletters}
\hspace{9.1cm}
\rule{-2mm}{0.1mm}\rule{8.7cm}{0.1mm}
\begin{multicols}{2}
\narrowtext\noindent
where $G^{\left(\nu\right)}_{ab}$, $G^{\left(\ell\right)}_{ab}, G^{\left(P\right)}_{ba}$, $G^{\left(U\right)}_{1b}$, $F^{\left(U\right)}_{\alpha b}$, $G^{\left(D\right)}_{1b}$, $F^{\left(D\right)}_{\alpha b}$, $G^{\left(J\right)}$ and $F^{\left(J\right)}_{\alpha\beta}$ are the Yukawa coupling constants. $\eta^*$, $\rho^*$ and $\chi^*$ denote the $\eta$, $\rho$ and $\chi$ antiparticle fields, respectively. \par 
Our goal in this Letter is to show that the 3-3-1 model furnishes a good candidate for (self-interacting) dark matter. The main properties that a good dark matter candidate must satisfy are stability and neutrality. Therefore, we go to the scalar sector of the model, more specifically to the neutral scalars, and we examine whether any of them can be stable and in addition whether they can satisfy the self-interacting dark matter criterions \cite{SS00}. In addition, one should notice that such dark matter particle must not overpopulate the Universe. On the other hand, since our dark matter particle is not imposed arbitrarily to solve this specific problem, we must check that the necessary values of the parameters do not spoil the other bounds of the model.\par
We can check through a direct calculation by employing the Lagrangians (\ref{pot}), (\ref{deri}) and (\ref{yukt}) and by using the eigenstates (\ref{3H0}) and (\ref{zetac}) that the Higgs scalar $h^0$ and $H^0_3$ can, in principle, satisfy the criterions above. Remarkably they do not interact directly with any standard model field except for the standard Higgs $H^0_1$. However, $h^0$ must be favored, since we have checked that it is easier to obtain a large scattering cross section for it, relative to $H^0_3$, by a convenient choice of the parameters. \par
In contrast to the singlet models of the Refs. \cite{MC02,BP01}, where an extra symmetry must be imposed to account the stability of the dark matter, here the decay of the $h^0$ scalar is automatically forbidden in all orders of perturbative expansion. This is because of the following features: i) this scalar comes from the triplet $\chi$, the one that induces the spontaneous symmetry breaking of the 3-3-1 model to the standard model. Therefore, the standard model fermions and the standard gauge bosons cannot couple with $h^0$, ii) the $h^0$ scalar comes from the imaginary part of the Higgs triplet $\chi$. As we mentioned above, the imaginary parts of $\eta$ and $\rho$ are pure massless Goldstone bosons. Therefore, there is not physical scalar fields which can mix with $h^0$. So, the only interactions of $h^0$ come from the scalar potential and they are $H_3^0h^0h^0$ and $H_1^0h^0h^0$. This latter has strength $2i\left(a_5v^2 + a_6u^2\right)/v_W \equiv 2i\Theta$. We can check also that $h^0$ does not interact with other exotic particle. \par
Hence, if $v \sim u \sim (100 - 200)$ GeV and $-1 \leq a_5 \sim a_6 \leq 1$, the $h^0$ can interact only weakly with ordinary matter through the Higgs boson of the standard model $H_1^0$. The relevant quartic interaction for scattering is $h^0h^0h^0h^0$, whose strength is $-ia_3$. Other quartic interactions evolving $h^0$ and other neutral scalars are proportional to $1/w$ and so we neglect them. The cross section of the process $h^0h^0 \to h^0h^0$ {\it via} the quartic interaction is $\sigma = a_3^2/64\pi m_h^2$. The contribution of the trilinear interactions {\it via} $H_1^0$ and $H_3^0$ exchange are negligible. There is no other contribution to the process involving the exchange of vector or scalar bosons. A self-interacting dark matter candidate must have mean free path $\Lambda = 1/n\sigma$ in the range 1 kpc $< \Lambda <$ 1 Mpc, where $n = \rho/m_h$ is the number density of the $h^0$ scalar and $\rho$ is its density at the solar radius \cite{SS00}. Therefore, with $a_3 = -1$, $-0.208 \times 10^{-7} {\mbox{ GeV}} \leq f \leq -0.112 \times 10^{-6} {\mbox{ GeV}}$, $w = 1000$ GeV, $u = 195$ GeV and $\rho = 0.4$ GeV/cm$^3$, we obtain the required Spergel-Steinhardt bound, {\it i. e.}, $2 \times 10^3 {\mbox{ GeV}}^{-3} \leq \sigma/m_h \leq 3 \times 10^4 {\mbox{ GeV}}^{-3}$ \cite{SS00}.\par
With this set of parameter values, we see from Eq. (\ref{mh}) that 5.5 MeV $\leq m_h \leq$ 29 MeV. This means that our dark matter particle is non-relativistic in the decoupling era (decoupling temperature $\sim$ 1 eV) and, for a standard model Higgs boson mass $\sim 100$ GeV \cite{Gea00}, it is produced by a thermal equilibrium density of the standard Higgs scalar to $h^0h^0$ pairs \cite{MC02}. The density of the $h^0$ scalar from the $H^0_1$ decay can be obtained following the standard procedure, {\it i. e.}, we must solve the Boltzmann equation
\begin{equation}
\frac{dn_h}{dt} + 3Hn_h = \langle\Gamma_H\rangle n^{({\rm eq})}_H,
\label{bolt}\end{equation}
where $n_h$ is the number density of the $h^0$ scalar at the time $t$, $H$ is the Hubble expansion rate, 
\begin{equation}
\Gamma_H = \frac{\Theta^2}{4\pi E}
\end{equation} 
is the decay rate for the $H_1^0$ with energy $E$ and
\begin{equation}
n^{({\rm eq})}_H = \frac{1}{2\pi^2}\int^\infty_{m_1}{\frac{E\sqrt{E^2 - m_1^2}}{{\rm e}^{E/T} - 1}}
\end{equation}
is the thermal equilibrium density of the standard $H^0_1$ at temperature $T$ \cite{KT90}. We are using the condition that the temperature is less than the electroweak phase transition $T_{\rm EW} \geq 1.5m_1$ \cite{MC02}. The thermal average of the decay rate is given by
\begin{equation}
\langle\Gamma_H\rangle = \frac{\alpha \left(\Theta T\right)^2}{8\pi^3n_H^{({\rm eq})}}{\rm e}^{m_1/T},
\end{equation}
where $\alpha$ is an integration parameter that can be taken to be 1.87 \cite{MC02}. We define $\beta \equiv n_h/T^3$ and in the radiation-dominated  era we write the evolution equation (\ref{bolt}) as
\begin{equation}
\frac{d\beta}{dT} = -\frac{\langle\Gamma_H\rangle\beta^{({\rm eq})}}{KT^3} = -\frac{\alpha}{8\pi^3K{\rm e}^{m_1/T}}\left(\frac{\Theta}{T^2}\right)^2,
\end{equation}
where $K^2 = 4\pi^3g\left(T\right)/45m_{\rm Pl}^2$, $\beta^{({\rm eq})} = n^{({\rm eq})}_h/T^3$ is the $\beta$ parameter in the thermal equilibrium, $m_{\rm Pl} = 1.2 \times 10^{19}$ GeV is the Planck mass and $g\left(T\right) = g_B + 7g_F/8 = 136.25$ for the model of the Ref. \cite{PT93}. $g_B$ and $g_F$ are the relativistic bosonic and fermionic degrees of freedom, respectively. Here we are taking $T = m_1$ since this regime gives the larger contribution to $\beta$ \cite{MC02}. Hence,
\begin{equation}
\beta = \frac{\alpha \Theta^2}{4 \pi^3Km_1^3}.
\label{beta}\end{equation}
Now, the cosmic density of the $h^0$ scalar is
\begin{equation}
\Omega_h = 2g\left(T_\gamma\right)T^3_\gamma\frac{m_h\beta}{\rho_cg\left(T\right)},
\label{den}\end{equation}
where $T_\gamma = 2.4 \times 10^{-4}$ eV is the present photon temperature, $g\left(T_\gamma\right) = 2$ is the photon degree of freedom and $\rho_c = 7.5 \times 10^{-47}h^2$, with $h = 0.71$, being the critical density of the Universe. Let us take $m_h = 7.75$ MeV, $v = 174$ GeV, $a_5 =0.65$, $-a_6 = 0.38$ (actually in our calculations, we have used a better precision for $a_5$ and $a_6$) and $m_1 = 150$ GeV. Thus, from Eqs. (\ref{beta}) and (\ref{den}) we obtain $\Omega_h = 0.3$. Therefore, without imposing any new fields or symmetries, the 3-3-1 model possesses a scalar field that can satisfy all the properties required for the self-interacting dark matter and that does not overpopulate the Universe. \par
The candidate for self-interacting dark matter that we propose here differs from the singlet models of Refs. \cite{MC02,BP01} in an important point. As we have discussed above it comes from a gauge model proposed with another motivation that has an independent phenomenology. Therefore, the values of the parameters that we impose here must not spoil the preexisting bounds. We can obtain $m_1 \approx 150$ GeV from Eq. (\ref{h1}) with $a_1 = 1.2$ and $a_2 =0.36 $. From Eq. (\ref{h3}) we have $m_3 = 1$ TeV. On the other hand, one should notice that $m_h$ has a small value since $-f \sim \left(10^{-7} - 10^{-6}\right)$ GeV and $u \sim 195$ GeV. However, $h^0$ does not couple to the particles of the standard model for the Higgs boson. Thus, it evades the present accelerator limits. The constants $a_5$ and $a_6$ do not enter in the masses of the particles of the model and so, it is free in this work \cite{TO96}. \par
In conclusion, it is a remarkable fact that the 3-3-1 model has an option for self-interacting dark matter without the need of imposing any new symmetry to stabilize it. We have shown that the Spergel-Steinhardt bound for self-interacting dark matter \cite{SS00} can be realized in the 3-3-1 model with a reasonable choice of the values of the parameters. It is argued, in the context of the singlet models of Refs. \cite{MC02,BP01} that such a scalar field can have origin in a more fundamental theory such as GUT or supergravity. One should notice that our work suggests that we do not need to go to very high energy to access the origin of the dark matter. The 3-3-1 model, which can manifest itself at energies of the order of a few GeV or less, can provide a dark matter particle. \par
\bigskip
D. F. thanks the Coordena\c c\~ao de Aperfei\c coamento de Pessoal de N\'\i vel Superior and M. D. T. thanks the Funda\c c\~ao de Amparo \`a Pesquisa no Estado de S\~ao Paulo (Processo No. 99/07956-3) for full financial support. We are grateful to Dr. B. V. Carlson for reading the manuscript. We also thanks John McDonald for helpful conversations.

\end{multicols}

\begin{references}
\bibitem{KM01} For a recent review see S. Khalil e C. Mu\~noz, Contemp. Phys. 43 (2002) 51.
\bibitem{Aea99} J. N. Abdurashitov, {\it et al.} (SAGE Collaboration), Phys. Rev. C 60 (1999) 055801; Phys. Rev. Lett. 77 (1996) 4708; W. Hampel {\it et al.} (GALLEX Collaboration), Phys. Lett. B 477 (1999) 127; Y. Fukuda {\it et al.} (Super Kamiokande Collaboration), Phys. Rev. Lett. 82 (1999) 2644; {\it ibid} 81 (1998) 1562.
\bibitem{Nea01} C. B. Netterfield {\it et al.}, Boomerang experiments (Report No. {\tt astro-ph/0104460}); N. W. Halverson {\it et al.}, DASI experiment (Report No. {\tt astro-ph/0104489}); A. T. Lee {\it et al.}, MAXIMA experiment (Report No. {\tt astro-ph/0104459}); P. de Bernardis {\it et al.}, Nature, 404 (2000) 955; S. Perlmuter {\it et al.} Astrophys. J. 517 (1999) 565; E. E. de Falco, C. S. Kochanek and J. A. Mu\~noz, Astrophys. J. 494 (1998) 74;  A. G. Riess {\it et al.} Astron. J. 116 (1998) 1009.
\bibitem{GM00} S. Ghigna, B. Moore, F. Governato, G. Lake, T. Quinn and J. Stadel, Astrophys. J. 544 (2000) 616; J. F. Navarro, C. S. Frenk, and S. D. M. White {\it ibid} 462 (1996) 563; B. Moore, T. Quinn, F. Governato, J. Stadel and G. Lake, Mon. Not. R. Astron. Soc. 310 (1999) 1147.
\bibitem{MG99} B. Moore, S. Ghigna, F. Governato, G. Lake, T. Quinn, J. Stadel and P. Tozzi, Astrophys. J. 524 (1999) L19; A. Klypin, A. V. Kravtsov, O. Valenzuela and F. Prada, Astrophys. J. 522 (1999) 82. 
\bibitem{DS01} R. Dav\'e, D. N. Spergel, P. J. Steinhardt and B. D. Wandelt, Astrophys. J., 547 (2001) 574.
\bibitem{BK00} J. S. Bullock, A. V. Kravtsov and D. H. Weinberg, Astrophys. J. 539 (2000) 517; R. A. Swaters, B. F. Madore and M. Trewhella, Astrophys. J. 531 (2000) L107; F. C. van den Bosch, B. E. Robertson, J. J. Dalcanton and W. J. G. de Blok, Astrophys. J. 119 (2000) 1579; J. F. Navarro and M. Steinmetz, Astrophys. J. 528 (2000) 607; C. Firmani, E. D'Onghia, V. Avila-Reese, G. Chincarini and X. Hernandez, Mon. Not. R. Astron. Soc. 315 (2000) L29; V. P. de Battista and J. A. Sellwood, Astrophys. J. 493 (1998) L5.
\bibitem{SS00} D. N. Spergel and P. J. Steinhardt, Phys. Rev. Lett. 84 (2000) 3760.
\bibitem{Yea00} N. Yoshida {\it et al.}, Astrophys. J. 544 (2000) L87.
\bibitem{BJ02} D. A. Buote, T. E. Jeltema, C. R. Canizares and G. P. Garmire, {\it Chandra evidence for a flatened, triaxial dark matter halo in the elliptical galaxy NGC 720} (Report No. {\tt arXiv:astro-ph/0205469}, to be published in Astrophys. J.). 
\bibitem{LB02} A. D. Lewis, D. A. Buote, J. T. Stocke, {\it Chandra observations of Abell 2029: the dark matter profile at $<$ 0.01R$_{\rm vir}$ in an unusually relaxed cluster} (Report No. {\tt arXiv:astro-ph/0209205}, submitted to the Astrophys. J.).
\bibitem{KK00} M. Kaplinghat, L. Knox and M. S. Turner, Phys. Rev. Lett. 85 (2000) 3335;  J. R. Primack, {\it The nature of dark matter}, Lectures at International School of Space Science, L'Aquila, Italy, 2001 (Report No. {\tt hep-ph/0112255}).
\bibitem{MC02} J. McDonald, Phys. Rev. Lett. 88 (2002) 091304.
\bibitem{BP01} C. P. Burgess, M. Pospelov and T. ter Veldhuis, Nucl. Phys. B 619 (2001) 709; M. C. Bento, O. Bertolami, R. Rosenfeld and L. Teodoro, Phys. Rev. D 62 (2000) 041302; M. C. Bento, O. Bertolami and R. Rosenfeld, Phys. Lett. B 518 (2000) 276; D. E. Holz and A. Zee, Phys. Lett. B 517 (2001) 239; V. Silveira and A. Zee, Phys. Lett. B 161 (1985) 136.  
\bibitem{FH93}  R. Foot, O. F. Hernandez, F. Pisano and V. Pleitez, Phys. Rev. D 47 (1993) 4158; F. Pisano and V. Pleitez, Phys. Rev. D 46 (1992) 410; P. H. Frampton, Phys. Rev. Lett. 69 (1992) 2889.
\bibitem{CQ99} For recent phenomenological works see J. E. Cieza Montalvo and M. D. Tonasse, Nucl. Phys. B 623 (2002) 325; Y. A. Coutinho, P. P. Queir\'oz Filho and M. D. Tonasse, Phys. Rev. D, 60 (1999) 115001; F. Cuypers, and S. Davidson, Eur. Phys. J. C 2 (1998) 503. 
\bibitem{TO02} M. D. Tonasse, Nucl. Phys. B 623 (2002) 316; H. N. Long, Phys. Rev. D 54 (1996) 4691; M. \"Ozer, Phys. Rev. D 54 (1996) 1143; J. C. Montero, F. Pisano and V. Pleitez, Phys. Rev. D 47 (1993) 2918.
\bibitem{PT93} V. Pleitez and M. D. Tonasse, Phys. Rev. D 48 (1993) 2353.
\bibitem{TO96} M. D. Tonasse, Phys. Lett. B 381 (1996) 191 [See also N. T. Anh, N. A. Ky and H. N. Long, Int. J. Mod. Phys. A 16 (2001) 541 and references cited therein].
\bibitem{Gea00} K. Hagiwara {\it et al.} (Particle Data Group), Phys. Rev. D 66 (2002) 010001-1.
\bibitem{KT90} See, for example, E. W. Kolb and M. S. Turner, {\it The early universe}, (Addison-Wesley Publishing Co., Reading, 1990).
\end{references}
\end{document}